%% file: main.tex
\documentclass[a4paper]{eccomas_2022}
\usepackage[utf8]{inputenc}
\usepackage[pdftex]{graphicx}
\usepackage{hhline}
\graphicspath{ {./images/} }
\usepackage{caption}
\usepackage{subcaption}
\usepackage{physics}
\usepackage{amsfonts}
\usepackage{algorithm}
\usepackage{pgf}
\usepackage{algpseudocode}
\usepackage[backend=bibtex]{biblatex}   % bibliography
\bibliography{mybib}
\usepackage{layouts}
\usepackage{pgfplots}
\usepackage[super]{nth}
\usepackage{enumitem}

\title{Machine Learning model for gas-liquid interface reconstruction in CFD numerical simulations}
\author{T. Nakano$^{1}$, M. A. Bucci$^{1}$, J-M. Gratien$^{2}$, T. Faney$^{2}$, G. Charpiat$^{1}$}

\address{$^{1}$ INRIA, LISN, bât 660, Universit\'e Paris-Saclay, 91405 Orsay cedex, France
\and
$^{2}$ IFPEN, 1 et 4, avenue du Bois-Pr\'eau, 92500 Rueil-Malmaison Cedex, France \url{http://www.ifpenergiesnouvelles.fr}}

\keywords{Graph neural network, interface reconstruction, VoF method}
\abstract{
The volume of fluid (VoF) method is widely used in multi-phase flow simulations to track and locate the interface between two immiscible fluids. A major bottleneck of the VoF method is the interface reconstruction step due to its high computational cost and low accuracy on unstructured grids. We propose a machine learning enhanced VoF method based on Graph Neural Networks (GNN) to accelerate the interface reconstruction on general unstructured meshes. We first develop a methodology to generate a synthetic dataset based on paraboloid surfaces discretized on unstructured meshes. We then train a GNN based model and perform generalization tests. Our results demonstrate the efficiency of a GNN based approach for interface reconstruction in multi-phase flow simulations in the industrial context.
}

\begin{document}

%\maketitle

%\tableofcontents

\input{sections/intro}

\input{sections/vof}

\input{sections/gnn-short}

\input{sections/dataset-eccomas}

\input{sections/training-eccomas}

\input{sections/validation-eccomas}

\input{sections/conclusion-eccomas}

\section{Acknowledgements}
This research was supported by DATAIA convergence Institute as part of the ``Programme d’Investissement d’Avenir”, (ANR- 17-CONV-0003) operated by INRIA and IFPEN.

\printbibliography
\end{document}

%% file: sections/intro.tex
\section{Introduction}
\label{sec:intro}
Gas-liquid inter-facial flow can be found in a large variety of industrial problems, such as cooling systems for electrical engines, chemical reactors or pore-scale flow in porous media. Among the many numerical methods developed to simulate such gas-liquid flows, an important issue is to track the motion of the free interface. Two classes of methods have emerged in the last decades: front tracking methods and interface capturing methods. The second class is more appropriate for industrial applications as it can more easily deal with topological changes of the free interface (e.g. break up or coalescence) in complex flows. The volume of fluid (VoF) method (\cite{kothe1996volume}, \cite{lorstad2004assessment}) is such an interface capturing method that recovers interface properties (i.e., normal, location, curvature) from the volume fraction of each fluid. One bottleneck of this method is the cost of the accurate computation of the local interface curvature which is essential for the evaluation of the surface tension force at the interface. Currently, the convolution method (\cite{williams2002numerical}, \cite{williams1998accuracy}) and the height function method (\cite{cummins2005estimating}, \cite{lorstad2004assessment}) are the most standard and the most accurate approaches. Nevertheless, these methods are computational expensive and while robust on regular structured grids, they are the source of unphysical fluid motions (\cite{harvie2006}) when applied to unstructured tetrahedral grids.

Several attempts to apply machine learning approaches have been done recently. Qi et al. \cite{qi2019computing} proposed a two-layer neural network to predict the curvature of a 2D surface, trained with a synthetic data-set. The trained model was then implemented into a CFD solver and tested on an unseen data-set. The prediction was accurate enough to consider this approach as a valid solution. Patel et al. \cite{patel2019computing} extended this approach to 3D geometry using a two-layer neural network as well. Their model was trained considering a synthetic data-set generated from a spherical surface. They then studied the performance of their trained model on analytical and CFD based test cases. Their approach outperformed the convolution method and even matched the accuracy of the height function method \cite{patel2019computing}. Svyetlichnyy et al. \cite{svyetlichnyy2018neural} also applied the neural network approach to predict the interface properties for a 2D and 3D surface reconstruction. Their model showed a good performance on the prediction of the normal but not on the curvature and surface location predictions. The major bottleneck of the common VoF methods and the proposed data-driven approaches is the low accuracy on unstructured grids. The VoF performs well on structured-like grids while it becomes unstable on an unstructured grid inducing artificial numerical residual currents. On the other hand  the proposed neural network solutions in Qi et al. \cite{qi2019computing}, Patel et al. \cite{patel2019computing} and Svyetlichnyy et al. \cite{svyetlichnyy2018neural} cannot be employed with unstructured grids since they do not take into account the variability in the neighbouring cells for a given mesh element. Below is a tabular summary of the involved problems for each available method today. It is worth noting that in industrial problems involving  multi-phase-flow simulations, unstructured grids are usually employed. To overcome this problem, we propose to use Graph Neural Networks (GNN) models to deal with complex data structures such as non-euclidean or graph-based inputs. 

\begin{table}[H]
   \centering
   \begin{tabular}{ l || c | r | r  }
     & Structured grid & Unstructured grid & Computational cost \\
     \hline \hline
     VoF & Good & Instability & Expensive \\ \hline
     Neural Network & Good & Not available & Less expensive \\
     \hline
   \end{tabular}
 \end{table}
 
The objective of this work is to propose a machine learning-based framework in place of conventional algorithms in order to model interface dynamics. This is expected to accelerate the interface reconstruction while preserving an accurate prediction. We start by presenting in section \ref{sec:vof} a quick review on the VoF method. We then present in section \ref{sec:gnn} a GNN architecture to recover interface properties in each mesh element from the discretized concentration field. In section \ref{sec:dataset}, we describe how to generate a training dataset that will allow the model to make accurate predictions on a large range of unseen test data representative of industrial applications. Section \ref{sec:Model_Training} details the model training procedure on the generated dataset. Finally, we validate our methodology in section \ref{sec:model_validation} using several synthetic test cases.

%% file: sections/vof.tex
\section{Volume of Fluid method}
\label{sec:vof}
There are a wide variety of numerical methods to solve multi-phase flow problems. Most of them are based on the resolution of the mass and momentum conservation equations leading to the Navier-Stokes (NS) equations. If we consider for instance two fluids A and B partitioning the simulation domain $\Omega$, a standard approach consists in representing each type of fluid as a single fluid with spatial physical properties and with the characteristic function $\chi_A(x)$ of fluid A, defined as follows: $\chi(x) = 1  \textrm{ if } x  \textrm{ in fluid A, else } \chi(x) = 0$. The interface between the two fluids induces a surface tension force that is modeled as a source term in the NS equation. The methods dealing with fluid interfaces differ in the way such interface is represented and computed. Among the various methods to solve such problems, the VoF method is an Eulerian-Eulerian approach that is capable of accurately capturing the interface from the phase indicator field $\alpha=\int_{\tau} \chi_A$ where $\tau$ is a cell of a mesh discretization $\Omega_h$ of $\Omega$. $\alpha$ represents then for each cell $\tau$ the volume fraction of fluid A. Fig.\ref{fig:bubble1} illustrates a bubble defined by a characteristic function $\chi_A$ in blue and the discretization on a mesh with cells colored depending on the value of the discrete phase indicator $\alpha$. Physical quantities such as the fluid density, viscosity and velocity are expressed as volume fraction weighted sums. For instance, let $\rho_A$ and $\rho_B$ be the density of fluid A and fluid B, the fluid density is then defined as $\rho = \chi_A \rho_A + (1-\chi_A) \rho_B$. The Navier-Stokes momentum and mass conservation equations can be written as:
$$\frac{\partial \rho \mathbf{U}}{\partial t}+\nabla \cdot(\rho \mathbf{U} \mathbf{U})=-\nabla p+\rho \mathbf{g}+\nabla \cdot(\mu \nabla \mathbf{U})+\rho \mathbf{S}+\mathbf{F_{\sigma}}$$
$$\frac{\partial \rho}{\partial t}+\nabla \cdot(\rho \mathbf{U})=0$$
where $t$, $\mathbf{U}$, $p$, $\rho$ and $\mu$ are time, the velocity, the pressure, the density and the dynamic viscosity. $\mathbf{g}$ represents body accelerations acting on the fluid, for example gravity. $\mathbf{F_{\sigma}}$ represents the surface tension force. The surface tension force term at the interface of two fluids is usually modeled as follows:
$$\mathbf{F_{\sigma}}=\sigma H \mathbf{n} \delta \left(x-x_{s}\right)$$
where $x_s$ are the points on the interface and $\sigma$, $H$ and $\mathbf{n}$ are the surface tension coefficient, the mean curvature of the interface and the normal direction to the interface. 
$\delta$ is the Kronecker-$\delta$ that allows to account for the surface tension only for points belonging to the interface. An extra advection equation is written to compute the evolution of $\chi_A$ discretized on the mesh with $\alpha$ the volume fraction.
\begin{figure}
    \centering
     \begin{subfigure}[b]{0.64\textwidth}
         \centering
         \includegraphics[width=1.0\textwidth]{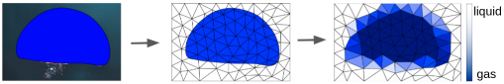}
         \caption{Discretization of continuous interfaces}
         \label{fig:bubble1}
     \end{subfigure}
     \hfill
     \begin{subfigure}[b]{0.35\textwidth}
         \centering
         \includegraphics[width=1.\textwidth]{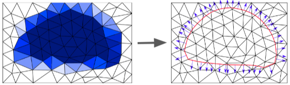}
         \caption{Interface reconstruction}
         \label{fig:bubble2}
     \end{subfigure}
    \caption{Gas-liquid interface representation with volume fraction}
    \label{fig:bubble1}         
\end{figure}

%% file: sections/gnn-short.tex
\section{Graph Neural Network Model}
\label{sec:gnn}
\subsection{GNN models}
Neural Networks (NN) are the most popular method used to recover data driven models. The most basic neural network, namely the Multi Layer Perceptron (MLP), is composed of a few layers of neurons (input, hidden and output layers). Data is fed to the input layer and predictions are made at the output layer. Neurons are connected between successive layers with multiplicative weights and non-linear activation functions. Convolutional Neural Networks (CNN) perform convolution operations with parameterized filters in order to capture spatial patterns on structured data on regular grids, such as pixel or voxel grids for images and videos. MLPs and CNNs cannot be applied to irregularly-organized data such as irregular meshes, which can be represented as graphs.

A graph consists of nodes and edges connecting the nodes as illustrated in Fig.\ref{fig:schema_graph}. We denote a graph by $\mathcal{G}=(\mathcal{V}, \mathcal{E})$ where $\mathcal{V}$ is a set of $n \in \mathbb{N}$ nodes $v_{i}$ and $\mathcal{E}$ is a set of $(e_{ij})_{i,j \in n }$ edges. $e_{ij}$ defines the connection between nodes $v_i$ and $v_j$. The graph connectivity can be represented by the adjacency matrix $\mathbf{A}$ which is a $n\cross n$ matrix with $A_{ij}=1$ if $e_{ij} \in \mathcal{E}$ and $A_{ij}=0$ if $e_{ij} \notin \mathcal{E}$. Nodes and edges can be equipped with features vectors $\mathbf{x}_v \in  \mathbb{R}^{d}$ (see Fig.\ref{fig:schema_graph}) and $\mathbf{x}_{e} \in \mathbb{R}^{m}$ respectively. Likewise, $\mathbf{X}_v\in \mathbb{R}^{n \cross d}$ is the node feature matrix, $\mathbf{X}_e\in \mathbb{R}^{n \cross m}$ is the edge feature matrix.

Graph Neural Networks (GNNs) are designed to treat data in the form of graph structures. A GNN takes a graph as its input and returns the same graph with updated features as the output. \textit{Message passing} (MP) is the main mechanism that allows the nodes to communicate through connections in order to update the node features. The MP mechanism on the $i$ node can be summarised in the following three steps:
\begin{enumerate}
    \item message computation: $\phi_v (\mathbf{x}_{v_j}) \rightarrow \tilde{\mathbf{x}}_{v_j}$ for all nodes $j$ in the neighbour of the $i$ node
    \item message propagation: $\phi_e (\tilde{\mathbf{x}}_{v_j}, \mathbf{x}_{e_{ij}}) \rightarrow \bar{\mathbf{x}}_{v_j}$
    \item  message aggregation: $\phi_a (\mathbf{x}_{v_i}, \underset{j}{\square}(\bar{\mathbf{x}}_{v_j})) \rightarrow \mathbf{x}_{v_i} $ where $\underset{j}{\square}$ is any aggregation function with permutation invariant properties. 
\end{enumerate}
$\phi_v$, $\phi_e$, $\phi_a$ are learnable functions, generally MLPs. The same steps are computed for all nodes in the graph in an operation similar to convolution. 
In our work we employ SAGEConv which is a variant of graph convolution network proposed by Hamilton et al.~\cite{Hamilton2017inductive}.
We refer the reader to \cite{wu2020comprehensive}, \cite{zhou2020graph} and \cite{Hamilton2017inductive} for a thorough review on the topic.

\subsection{GNN architecture design}

\subsubsection{Data modeling}
\begin{figure}
     \centering
     \begin{subfigure}[t]{0.28\textwidth}
         \centering
         \includegraphics[width=0.90\linewidth]{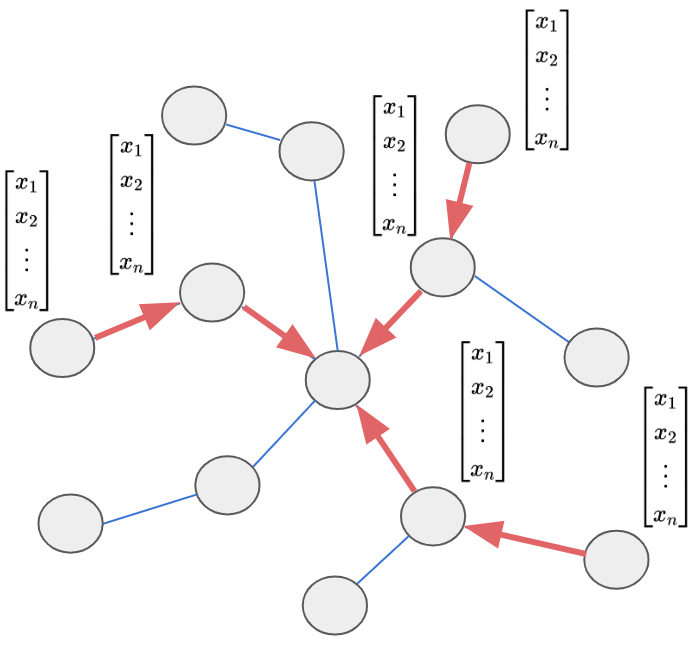}
         \caption{Schematic diagram of a graph}
    \label{fig:schema_graph}
     \end{subfigure}     
     \begin{subfigure}[t]{0.34\textwidth}
         \centering
         \includegraphics[width=0.90\linewidth]{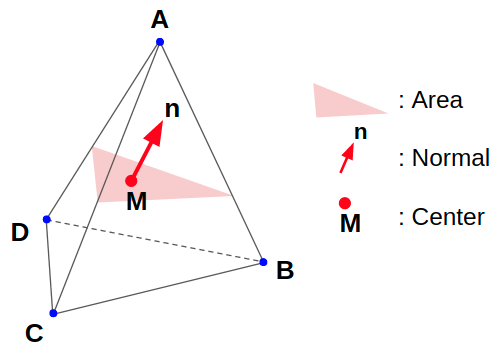}
         \caption{Surface properties}
         \label{fig:properties}
     \end{subfigure}
     \begin{subfigure}[t]{0.34\textwidth}
         \centering
         \includegraphics[width=0.90\linewidth]{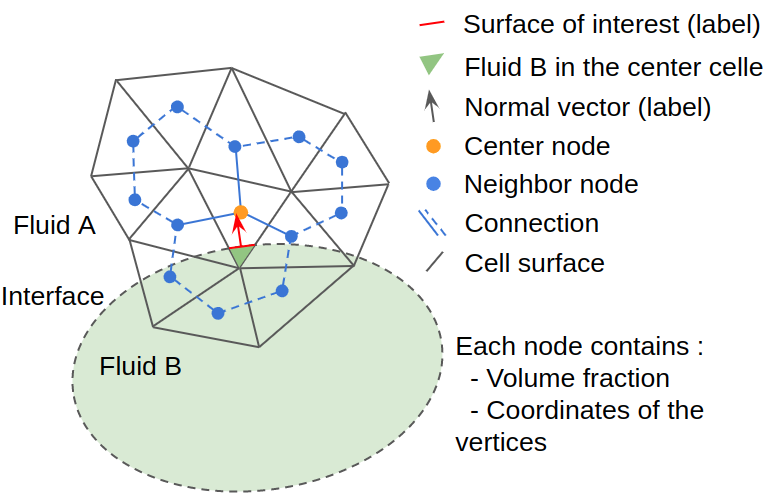}
         \caption{Graph structure}
         \label{fig:graph}     
     \end{subfigure}
     \caption{Gas-liquid interfaces representation with GNN graphs}
\end{figure}

The objective is to predict the normal, the curvature, the center and the area of the reconstructed surface for each tetrahedral cell from a graph containing the information from neighbouring tetrahedral cells (including the current cell). Therefore, each interface cell is associated to its own graph and interface properties, representing a single data point in the training process. The nodes of the graph are the centroids of each cell, and the edges are defined by the cell connectivity through shared faces. The dimension of the graph is defined by all elements sharing at least one vertex with the current cell. The inputs are provided to the graph as node features for each cell: coordinates of cell vertices and volume fraction. Fig.\ref{fig:properties} sketches the reconstructed surface properties for a three dimensional tetrahedral cell, and Fig.\ref{fig:graph} shows the graph for a given interface cell in two dimensions.

In the training algorithm, all the coordinates are normalized by $L = \textrm{max} (L_{x}, L_{y}, L_{z})$, where $L_{x}$, $L_{y}$ and $L_{z}$ are the maximum dimensions in the $x,y$ and $z$ directions over all mesh cells, such that each tetrahedral cell belongs to the unitary cube. The corresponding adimensional curvature $H$ is $H = H'L$ where $H'$ is the original curvature, and the corresponding adimensional area of the reconstructed surface $S$ is $S = S'/L^2$ where $S'$ is the original area. The coordinates of the interface center $M$ are expressed in the barycentric system as $M = p_1 P_1 + p_2 P_2 + p_3 P_3 + p_4 P_4$, where $P_1, P_2, P_3, P_4$ are the coordinates of each vertex of the tetrahedral cell (see Fig.\ref{fig:properties}). $p_1,p_2,p_3,p_4$ are the barycentric coordinates of $M$ and they satisfy the constraint $p_1 + p_2 + p_3 + p_4 = 1$.

\subsubsection{Neural network architecture}
\begin{figure}\centering
\subfloat[NN-1: Normal/Curvature]{\label{fig:nns1}\includegraphics[width=0.49\linewidth]{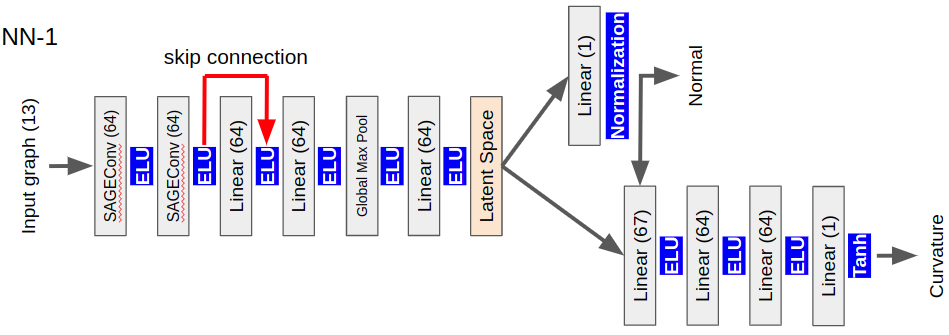}}\hfill
\subfloat[NN-2: Center]{\label{fig:nns2}\includegraphics[width=0.49\linewidth]{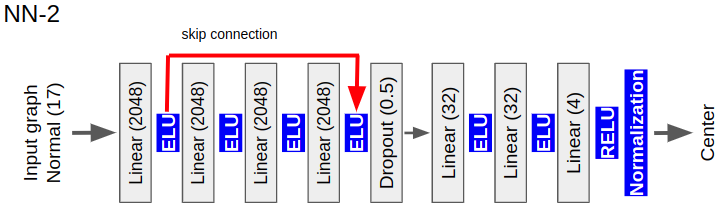}}\par
\subfloat[NN-3: Area]{\label{fig:nns3}\includegraphics[width=0.49\linewidth]{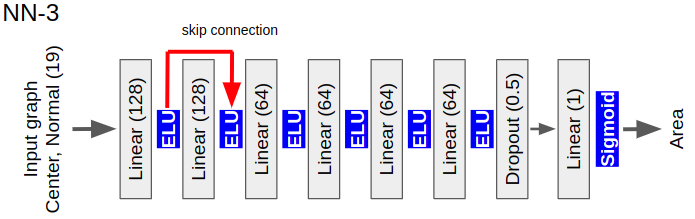}}
\caption{Schematic diagram of the architecture, the numbers in the figure mean the number of node in the layer or the probability of dropout}
\label{fig:architecture}
\end{figure}

Fig.\ref{fig:architecture} shows the architecture of our model. The design of the architecture is based on a causality criterion: first the normal and the curvature of the interface are computed, then the interface location is defined through the computation of the center by moving the interface in the computed normal direction. Finally, the area of the interface is evaluated as a function of the normal, the center and the vertices of the considered cell. The whole model then consists in staking three different sub-models. 

NN-1 is a graph neural network that simultaneously predicts the normal and the curvature from the input graph associated to a given interface cell. The node features are the coordinates of the cell vertices and the volume fraction $\alpha$ of the cells. More specifically, at each node we have the $3$ dimensional coordinates for the $4$ vertices of a tetrahedron $\{ \mathbf{P}_v \in \mathbb{R}^{12}, \forall v \in \mathcal{V} \}$. Here the coordinates of vertices are flattened in a manner that $\mathbf{P}_v = (P_1, P_2,..., P_4) = (x_1, y_1, z_1, x_2, y_2, \ldots, z_4)$ where $1,2,3,4$ are the number of the vertices of the tetrahedron. We then concatenate $\mathbf{P}_v$ with the volume fraction of the same node as $\mathbf{x}_v = \left( \alpha, \mathbf{P}_v \right) \in \mathbb{R}^{13}$. Therefore, an input feature of NN-1 for a graph having $n$ nodes can be denoted as $\mathbf{X}_{1} \in \mathbb{R}^{13 \cross n}$. The first two layers of NN-1 are SAGEConv layers that process the input graph and return a new graph with updated features. Node features are further processed by two linear layers. Skip connections are employed to prevent the vanishing gradient problem. The following global-mean-pool layer returns graph averaged node features. A latent representation of the initial graph is then recovered. The model splits into two branches of linear layers for the prediction of the normal $\mathbf{n} \in \mathbb{R}^{3}$ and the curvature $H$.

NN-2 is a simple MLP network that predicts the interface center $M$ from the local information available in the current interface cell, including the normal $\mathbf{n}$ predicted by NN-1. We concatenate the volume fraction $\alpha$, the coordinates of the vertices $\mathbf{P}_{cell} \in \mathbb{R}^{3 \cross 4}$ and the normal $\mathbf{n}$, which is fed in the form of the dot product with $\mathbf{P}_{cell}$, that is, $\mathbf{n} \cdot \mathbf{P}_{cell} \in \mathbb{R}^{4}$. This helps the neural network discriminate vertices that are above or below the interface. Finally, a concatenated form of them $\mathbf{X}_2 = \left( \alpha, \mathbf{n} \cdot \mathbf{P}_{cell}, \mathbf{\hat{P}}_{cell} \right) \in \mathbb{R}^{17}$ is used as input, where $\mathbf{\hat{P}}_{cell}=(P_{cell1}, P_{cell2},..., P_{cell4}) \in \mathbb{R}^{12}$ is the flattened $\mathbf{P}_{cell}$ and $P_{cell1}, P_{cell2},...$ are the coordinates of the vertices. At the output the barycentric coordinates $p \in \mathbb{R}^4$ are recovered. A final $L1$ normalisation is applied on the output to enforce $\sum_{j=1}^4 p_j=1$.

NN-3 is also a simple MLP that predicts the interface area from the available cell quantities $\mathbf{X}_3 = \left(\mathbf{n} \cdot \mathbf{P}_{cell}, \mathbf{\hat{P}}_{cell}, p \right) \in \mathbb{R}^{19}$. A sigmoid function is used after the final layer. Due to the non-dimensionalization by $L$ of the coordinates, the maximum possible area is $\sqrt{2}/2$ which is less than $1$, allowing us to skip normalization of the target area.

%% file: sections/dataset-eccomas.tex
\section{Dataset}
\label{sec:dataset}

\subsection{Dataset generation}
\label{subsec:Data Generation}
\begin{figure}
     \centering
     \begin{subfigure}[t]{0.4\textwidth}
         \centering
         \includegraphics[width=0.7\linewidth]{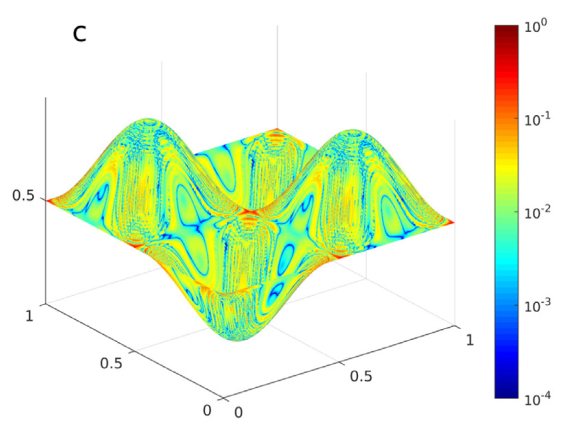}
     \end{subfigure}     \begin{subfigure}[t]{0.4\textwidth}
         \centering
         \includegraphics[width=0.7\linewidth]{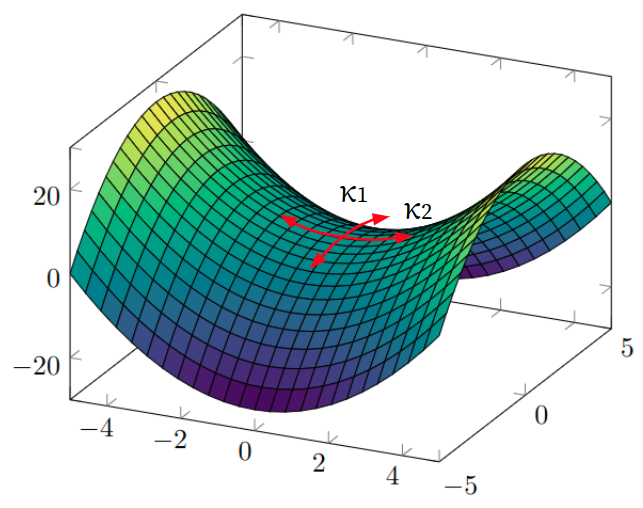}
     \end{subfigure}
     \caption{Curvature prediction by Patel et al.\cite{patel2019computing} (left) and Paraboloid (right)}
     \label{fig:sphare_parabo}
\end{figure}

Previous studies such as Patel et al. \cite{patel2019computing} uses a spherical surface to generate their synthetic dataset. They performed a prediction of the normal on a sinusoidal surface as shown in Fig.\ref{fig:sphare_parabo}. We observe that the largest errors are located on saddle point regions, where the surface presents discordant principal curvature sign. It is likely that the model trained on a  sphere based dataset is biased and can't generalize to surfaces with non-constant curvatures. To avoid this issue, we use a paraboloidal surface as shown in Fig.\ref{fig:sphare_parabo} to generate our dataset. A paraboloid can be described as $\frac{x^2\kappa_1}{2} + \frac{y^2\kappa_2}{2} = z$, where $\kappa_1$ and $\kappa_2$ correspond to the two curvatures along the principal planes. By sampling explicitly $\kappa_1$ and $\kappa_2$, we can obtain a dataset representative of a larger range of 2-dimensional surfaces in 3-dimensional space. Our dataset sampling process is as follows: \\

\begin{enumerate}[topsep=0.5pt, partopsep=0pt, itemsep=0pt, parsep=0pt]

\item Define the curvature range as $ \kappa_1, \kappa_2 \sim -[0.001, 0.5] \cup +[0.001, 0.5]$ and let $u_1 \sim \mathcal{U}(-1,1)$ and $u_2 \sim \mathcal{U}(0,1)$ be uniformly distributed random variables. This paraboloidal surface mimics the interface between two fluids.
\item Form a cubic space around the interface and discretize the inside with a tetrahedron mesh. The dimension of the cube is $(12+u_1) \times (12+u_1) \times (12+u_1)$, its center is given as $(u_1, u_1, u_1)$. The grid resolution is constantly $\Delta = 1$. The curvature in a dimensionless form will be: $\kappa_1\Delta, \kappa_2\Delta \sim \pm[0.001, 0.5]$.
\item Generate a graph around the center of the cube $(u_1, u_1, u_1)$ as illustrated in Fig.\ref{fig:graph}. As a reminder, we collect the following data from this graph. The label variables (normal, curvature, center and area) are collected at the center cell (see Fig.\ref{fig:graph} for its definition), and the feature variables (the coordinates of the vertices of a tetrahedron, edges, volume fraction) are collected at all nodes in the graph.
\item Rotate the graphs randomly by $(rot_x, rot_y, rot_z) = (\cos^{-1} (2u_2-1), \pi u_2, 2\pi u_2)$. Those formula are known to give a uniform 3D rotation.
\item Repeat this step from 1 to 5 for $N_{all}$ times, $N_{all}$ being the dimension of the desired dataset.
\end{enumerate}

\subsection{Dataset analysis}

\begin{figure}
     \centering
     \begin{subfigure}[t]{0.24\textwidth}
         \centering
         \includegraphics[width=0.98\textwidth]{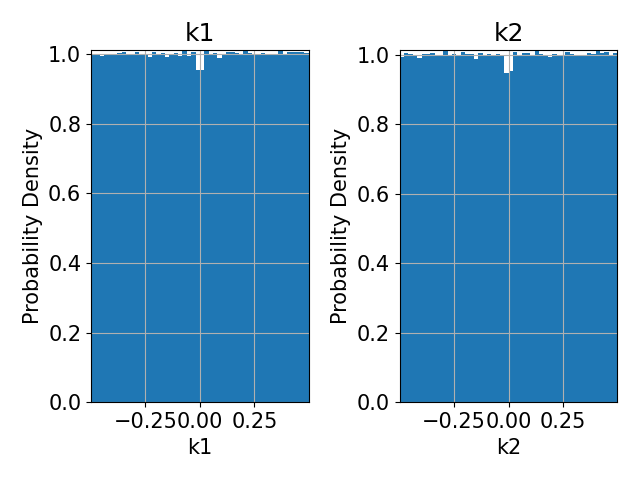}
     \end{subfigure}
     \hfill
     \begin{subfigure}[t]{0.24\textwidth}
         \centering
         \includegraphics[width=0.98\textwidth]{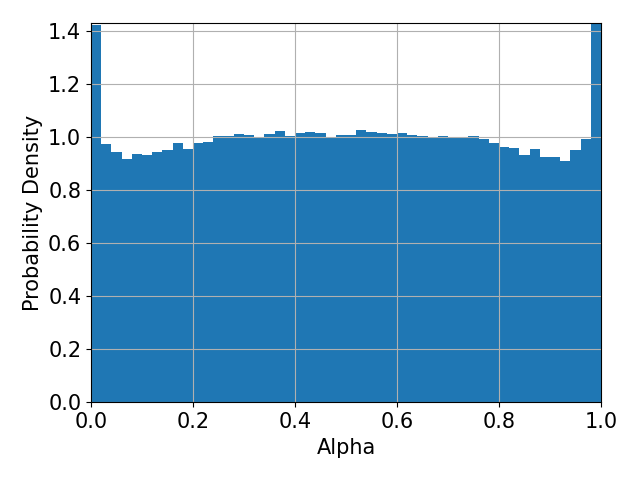}
     \end{subfigure}
     \hfill
     \begin{subfigure}[t]{0.24\textwidth}
         \centering
         \includegraphics[width=0.98\textwidth]{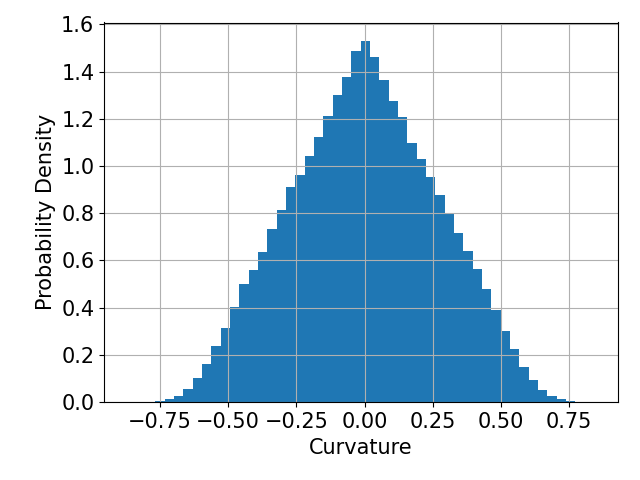}
     \end{subfigure}
     \hfill
     \begin{subfigure}[t]{0.24\textwidth}
         \centering
         \includegraphics[width=0.98\textwidth]{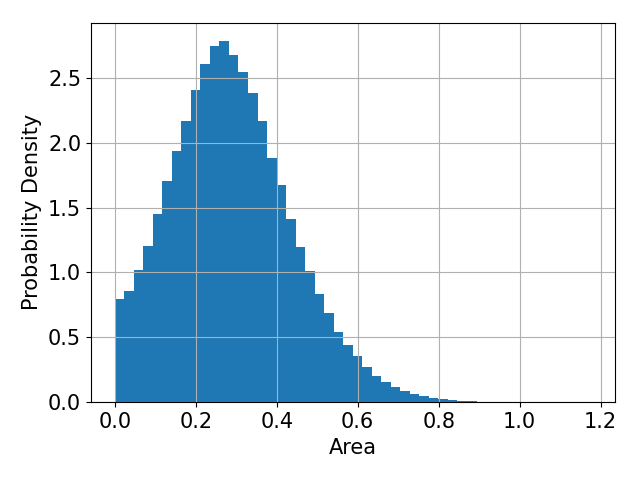}
     \end{subfigure}
     \caption{Histograms on the generated dataset: $\kappa_1$, $\kappa_2$, $\alpha$, Curvature and Area (left to right)}
     \label{fig:hists_gen_data}
\end{figure}

We generate $N_{all}=500,000$ paraboloids for the dataset. Fig.\ref{fig:hists_gen_data} shows the histograms of some properties of the generated dataset. Both $\kappa_1$ and $\kappa_2$ have a uniform distribution. The mean curvature is defined as $H=\frac{\kappa_1+\kappa_2}{2}$. The area and the mean curvature is non-uniform and non-controllable a priori. Nevertheless, we expect such a dataset generation procedure to better represent general surfaces than a dataset solely based on spherical surfaces.

%% file: sections/training-eccomas.tex
\section{Model training}
\label{sec:Model_Training}

We divide our 500,000 graphs dataset into a 7:2:1 ratio for training,  validation and test. To train more efficiently, some exact label values can also be used as inputs: in NN-2, the input normal can be the normal predicted by NN-1 or the exact normal. This means that two different loss functions need to be defined. The same is true for the the input center in NN-3. Each individual loss is a L2 normalized MSE loss. The total loss is then:
$$\mathrm{Loss_{total}} =\mathrm{Loss_{normal}} + \mathrm{Loss_{curv}} + \mathrm{Loss_{center}^{wl}} + \mathrm{Loss_{center}^{wol}} + + \mathrm{Loss_{area}^{wl}} + \mathrm{Loss_{area}^{wol}}$$
where $\mathrm{Loss_{normal}}$, $\mathrm{Loss_{curv}}$, $\mathrm{Loss_{center}}$ and $\mathrm{Loss_{area}}$ are the loss functions of the normal, curvature, center and area predictions. "wl" means that the label values are used for the input (wl: with-label). "wol" means that the predicted values are used for the input (wol: without-label).

The pytorch-integrated adaptive learning rate is used and set between $\mathrm{1e\mbox{-}2}$ and $\mathrm{1e\mbox{-}6}$, reducing it by a factor of $0.9$ when the validation loss function doesn't improve for $5$ consecutive epochs. The batch size is 512. The \textit{early stopping} \textit{patience} parameter  is set to $80$ epochs in order to avoid overfitting. Finally, gradient clipping is set to $\mathrm{1e\mbox{-}2}$ and the Adam optimizer is used.

\begin{figure}[H]\centering
\subfloat{\includegraphics[width=.32\textwidth]{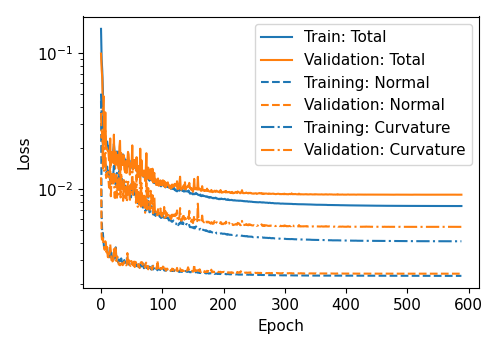}}
\subfloat{\includegraphics[width=.32\textwidth]{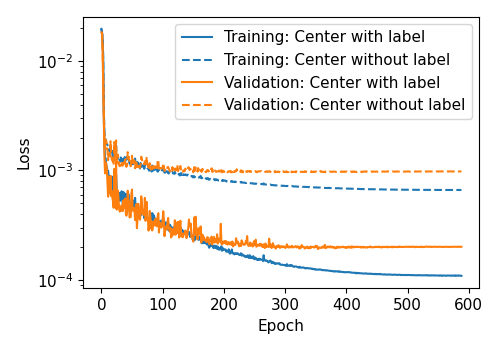}}
\subfloat{\includegraphics[width=.32\textwidth]{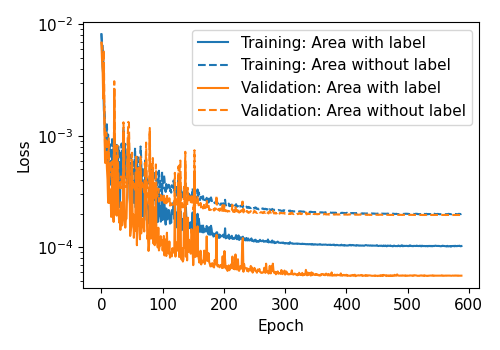}}
\caption{Learning curves: total, normal, curvature (left), center (middle) and area (right)}
\label{fig:learning_curves}
\end{figure}

Fig.\ref{fig:learning_curves} shows the learning curves of the total and individual train/validation losses during training. No improvement was observed for the total validation loss after \nth{508} epoch and the training was stopped at the \nth{588} epoch. All types of losses converge without major overfitting. For the center and the area, the loss with label gives a lower loss function than without. This is expected since the input variables in the without label case are predicted variables by another NN and their variables can already contain errors. Those total and individual loss functions at the \nth{508} epoch for the training and the validation are compared in the Table \ref{tab:gen_loss}. It also shows the prediction on the test dataset by the model trained above. All types of losses stay in the same range as the ones in the training and the validation. This result confirms the generalization of the model trained above. 

\begin{figure}\centering
\subfloat{\includegraphics[width=.25\linewidth]{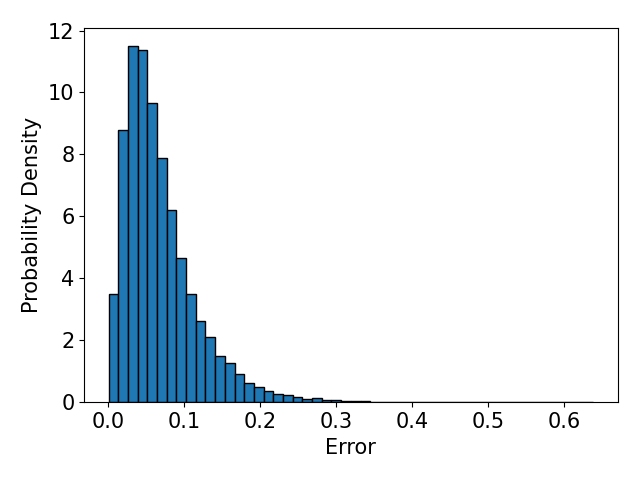}}
\subfloat{\includegraphics[width=.25\linewidth]{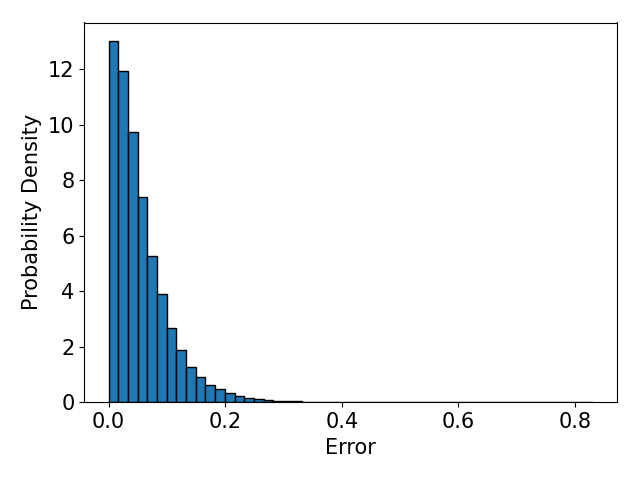}}
\subfloat{\includegraphics[width=.25\linewidth]{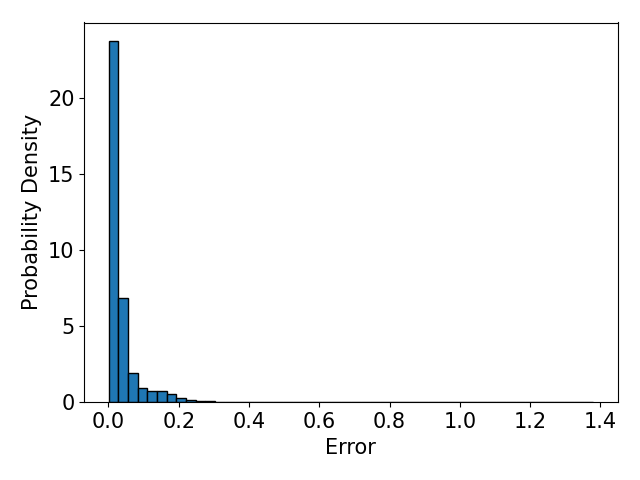}}
\subfloat{\includegraphics[width=.25\linewidth]{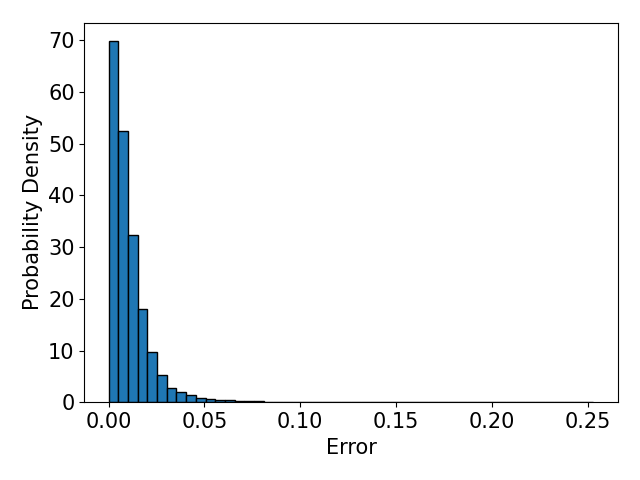}}
\caption{Histogram of the errors: Normal, Curvature, Center, Area}
\label{fig:e_hist_test1}
\end{figure}

To analyse the results in-depth, the errors in $L_1$ norm for each predicted property, as defined in Table \ref{tab:medi_errors_test1}, are plotted in Fig \ref{fig:e_hist_test1}. The histograms show a unimodal shape centered on zero with a long tail for large errors. Table \ref{tab:medi_errors_test1} also shows the median of the errors. Fig.\ref{fig:r2score_test1} shows the $R^2$ score between the prediction and their labels in the test dataset for the scalar variables, \textit{i.e.}, the curvature and the area. The scattered points are colored by the volume fraction $\alpha$ of the cell where the prediction was done. $R^2$ value is larger than $0.9$ for both of them. The value of $\alpha$ doesn't seem to have a clear relation with the error: high and low $\alpha$ values appear homogeneously in the figure regardless of wether the prediction under- or over- estimates the label.

%\noindent
\begin{minipage}[t]{0.6\textwidth}%
\centering
\captionof{table}{Model performance at the \nth{508} epoch}
\label{tab:gen_loss}
\begin{tabular}[t]{lccc}
\hline
Type of loss & Training & Validation & MSE\\
\hline
$\mathrm{Loss_{total}}$&$7.50e\mbox{-}3$&$9.07e\mbox{-}3$&$9.30e\mbox{-}3$ \\
$\mathrm{Loss_{normal}}$&$2.30e\mbox{-}3$&$2.39e\mbox{-}3$&$2.41e\mbox{-}3$ \\
$\mathrm{Loss_{curv}}$&$4.13e\mbox{-}3$&$5.26e\mbox{-}3$&$5.40e\mbox{-}3$ \\
$\mathrm{Loss_{center}^{wl}}$&$1.10e\mbox{-}4$&$2.01e\mbox{-}4$&$1.84e\mbox{-}4$ \\
$\mathrm{Loss_{center}^{wol}}$&$6.66e\mbox{-}4$&$9.76e\mbox{-}4$&$1.05e\mbox{-}3$ \\
$\mathrm{Loss_{area}^{wl}}$&$1.03e\mbox{-}4$&$5.53e\mbox{-}5$&$5.51e\mbox{-}5$ \\
$\mathrm{Loss_{area}^{wol}}$&$1.99e\mbox{-}4$&$1.95e\mbox{-}4$&$2.03e\mbox{-}4$ \\
\hline
\end{tabular}
\end{minipage}%
\begin{minipage}[t]{0.4\textwidth}%
\centering
\captionof{table}{Model performance on test set}
\label{tab:medi_errors_test1}
\begin{tabular}[t]{lccc}
\hline
Predicted variable&Median\\
\hline
$n_{error}=\norm{{\mathbf{n}_{label}-\mathbf{n}_{pred}}}$&$5.61e\mbox{-}2$\\
$H_{error}=\abs{H_{label}-H_{pred}}$&$4.16e\mbox{-}2$\\
$M_{error}=\norm{M_{label}-M_{pred}}$&$1.98e\mbox{-}2$\\
$A_{error}=\abs{A_{label}-A_{pred}}$&$7.57e\mbox{-}3$\\
\hline
\end{tabular}
\end{minipage}%

\bigskip

%% file: sections/validation-eccomas.tex
\section{Model validation}
\label{sec:model_validation}
\begin{figure}
\begin{minipage}[t]{.49\textwidth}\centering
\subfloat{\includegraphics[width=.5\textwidth]{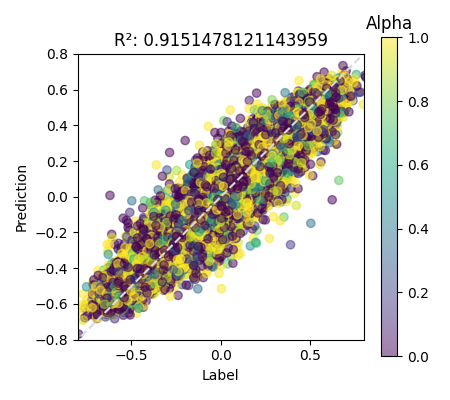}}
\subfloat{\includegraphics[width=.5\textwidth]{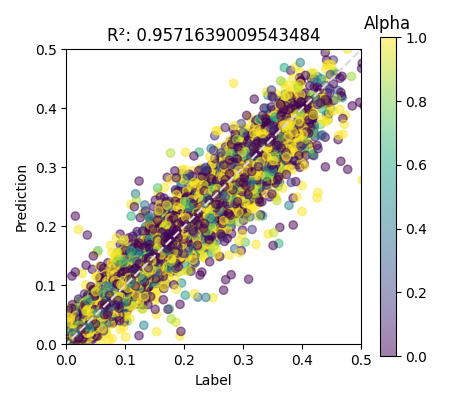}}
\caption{$R^2$ score on the prediction: Curvature (left) and Area (right)}
\label{fig:r2score_test1}
\end{minipage}
\hfill
\begin{minipage}[t]{.49\textwidth}\centering
\subfloat{\includegraphics[width=.5\textwidth]{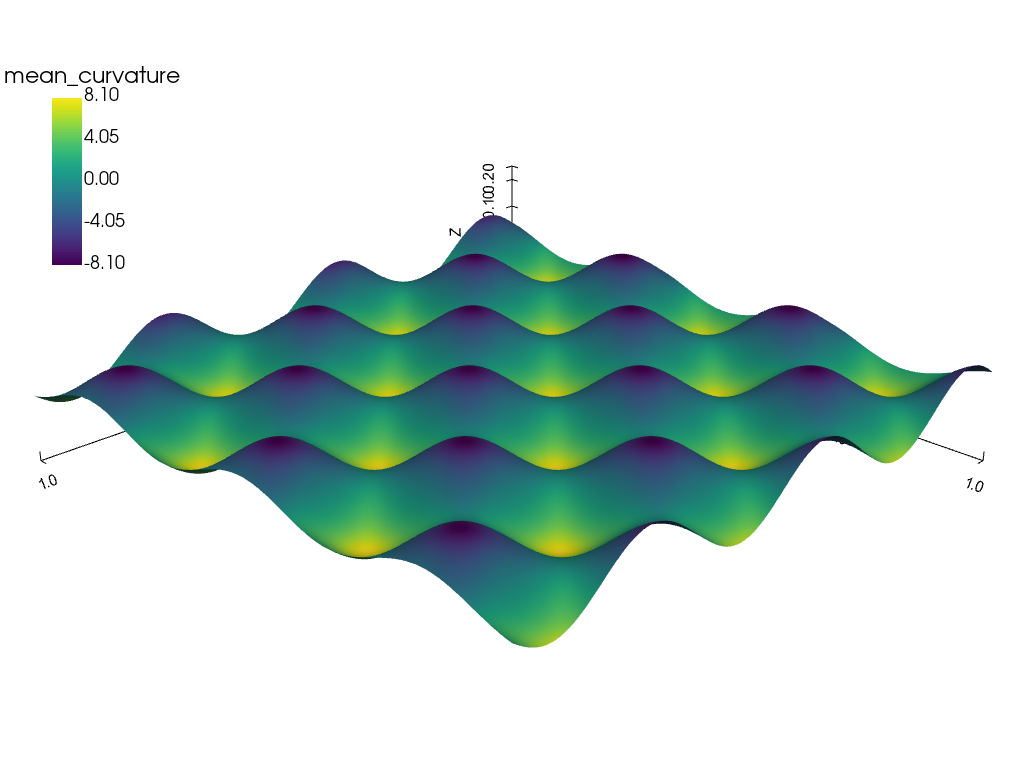}}
\subfloat{\includegraphics[width=.5\textwidth]{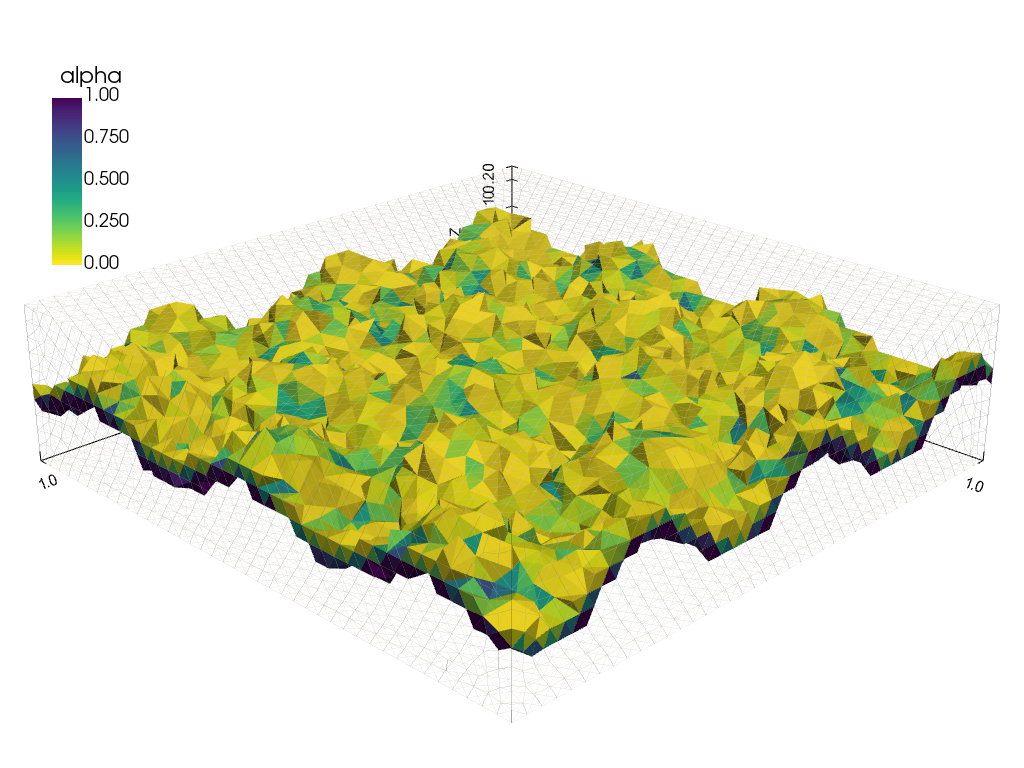}}
\caption{Test surface: colored by the mean curvature (left), voxelized and colored by $\alpha$ (right)}
\label{fig:test_surface}
\end{minipage}
\end{figure}

\begin{figure}\centering
    \subfloat{\includegraphics[width=.33\textwidth]{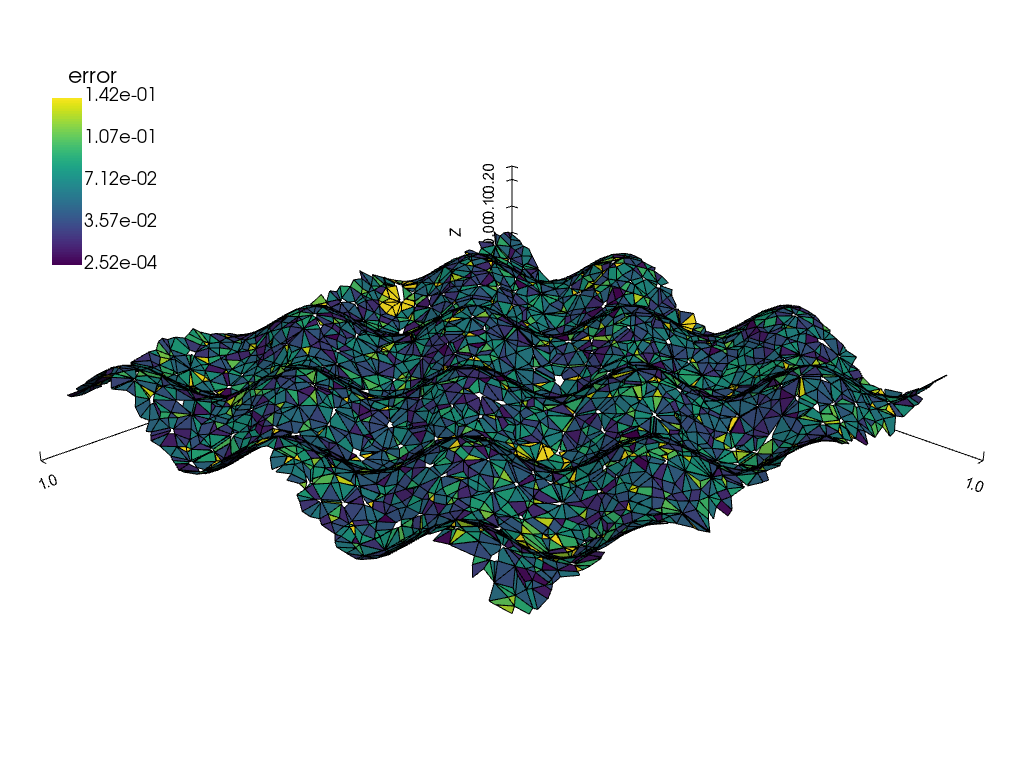}}
    \subfloat{\includegraphics[width=.33\textwidth]{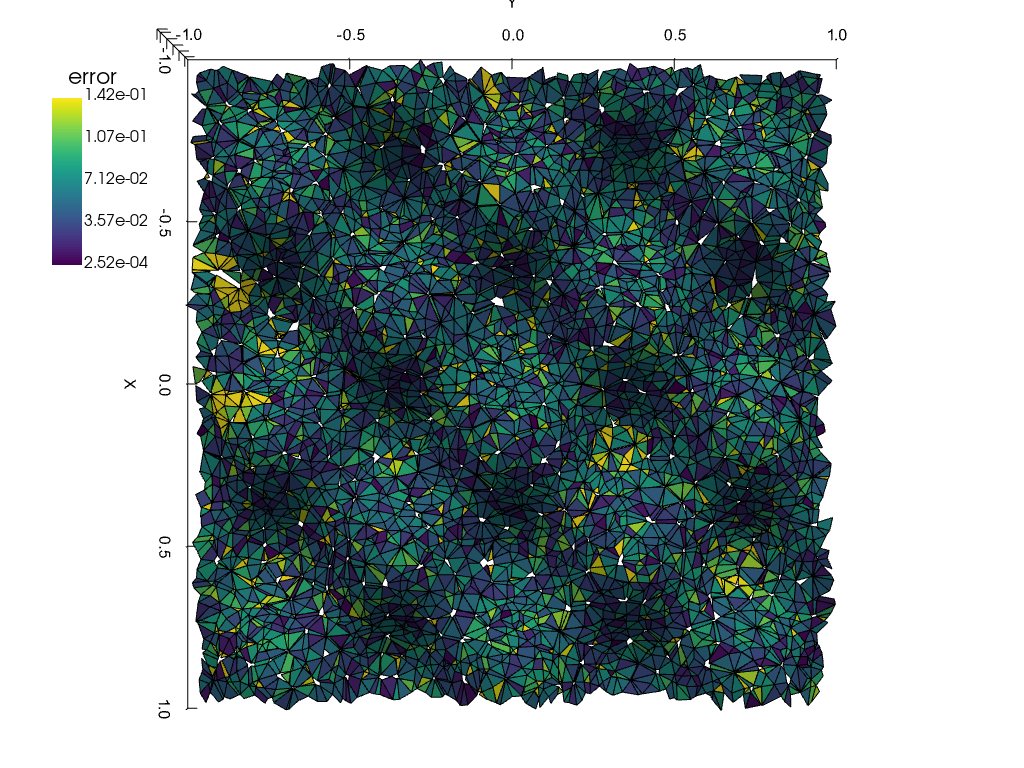}}
    \subfloat{\includegraphics[width=.33\textwidth]{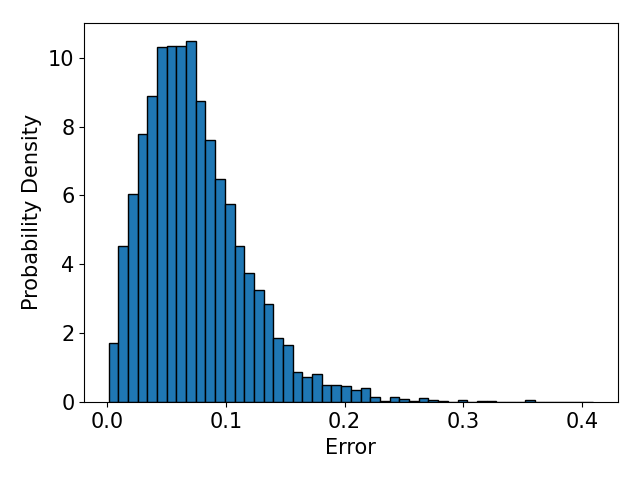}}
    \par
    \subfloat{\includegraphics[width=.33\textwidth]{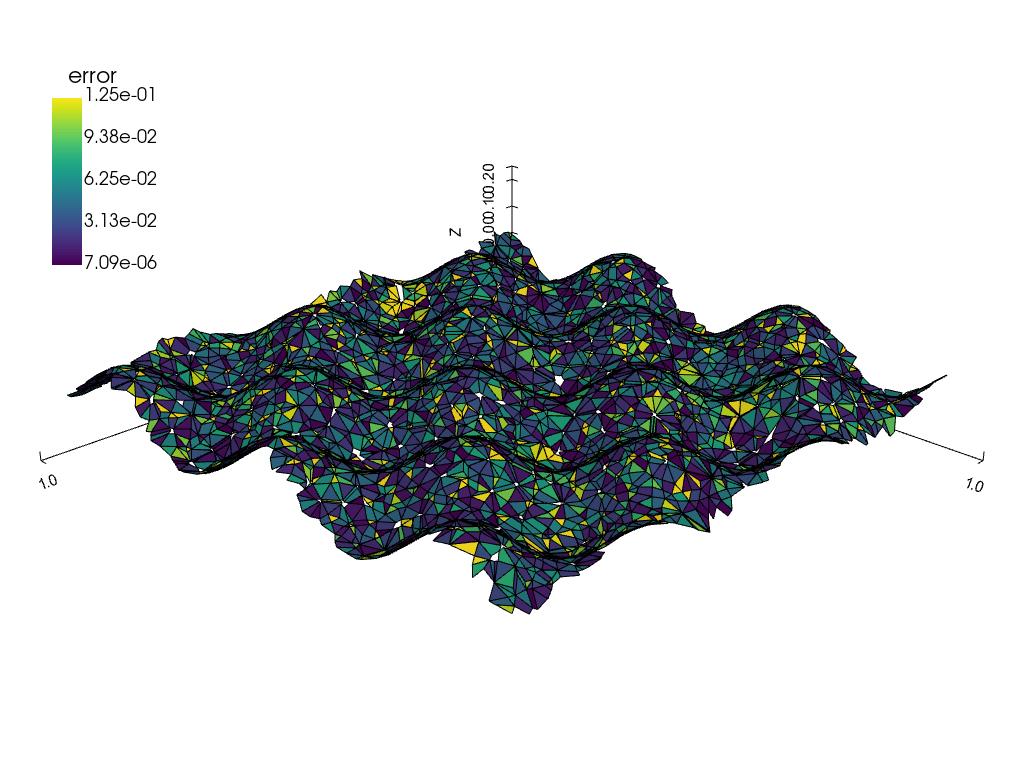}}
    \subfloat{\includegraphics[width=.33\textwidth]{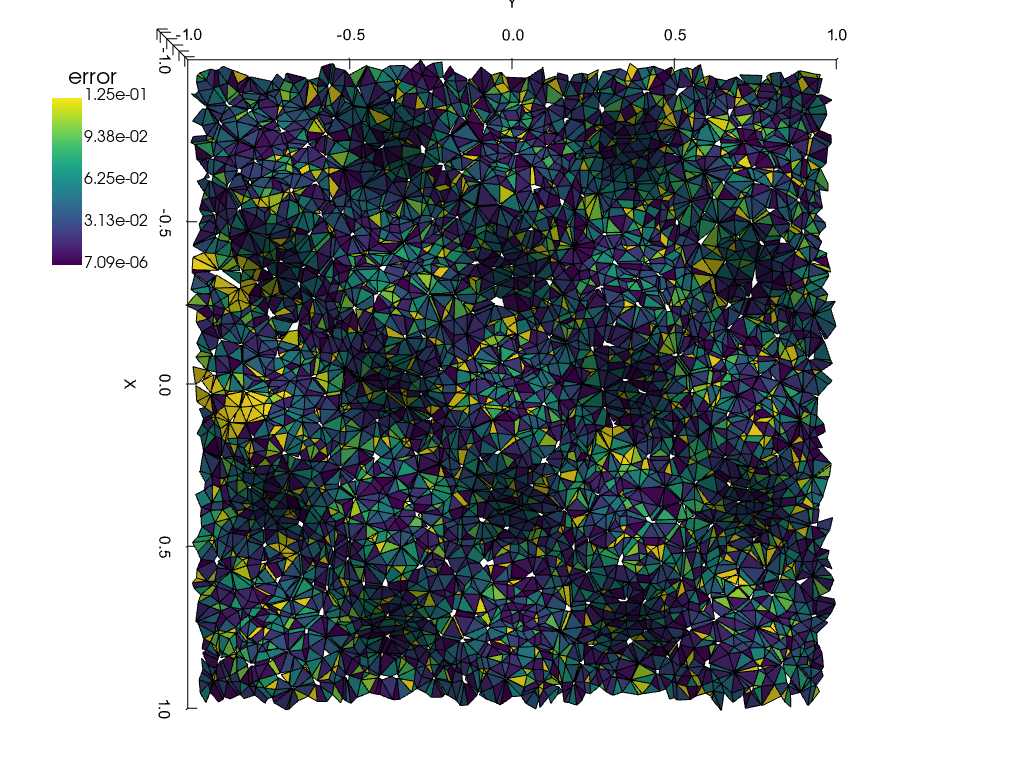}}
    \subfloat{\includegraphics[width=.33\textwidth]{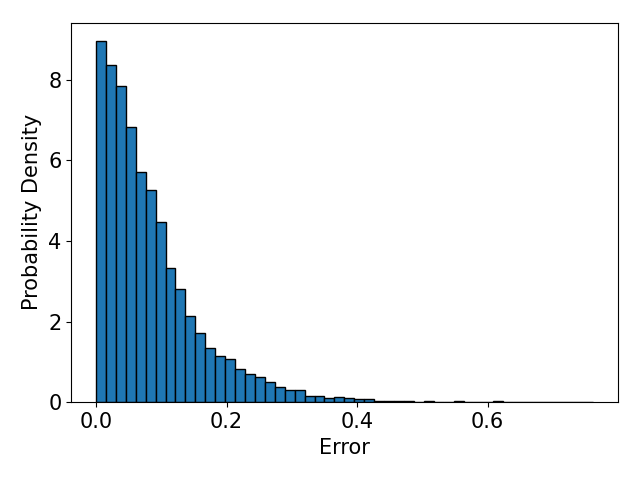}}
    \par
    \subfloat{\includegraphics[width=.33\textwidth]{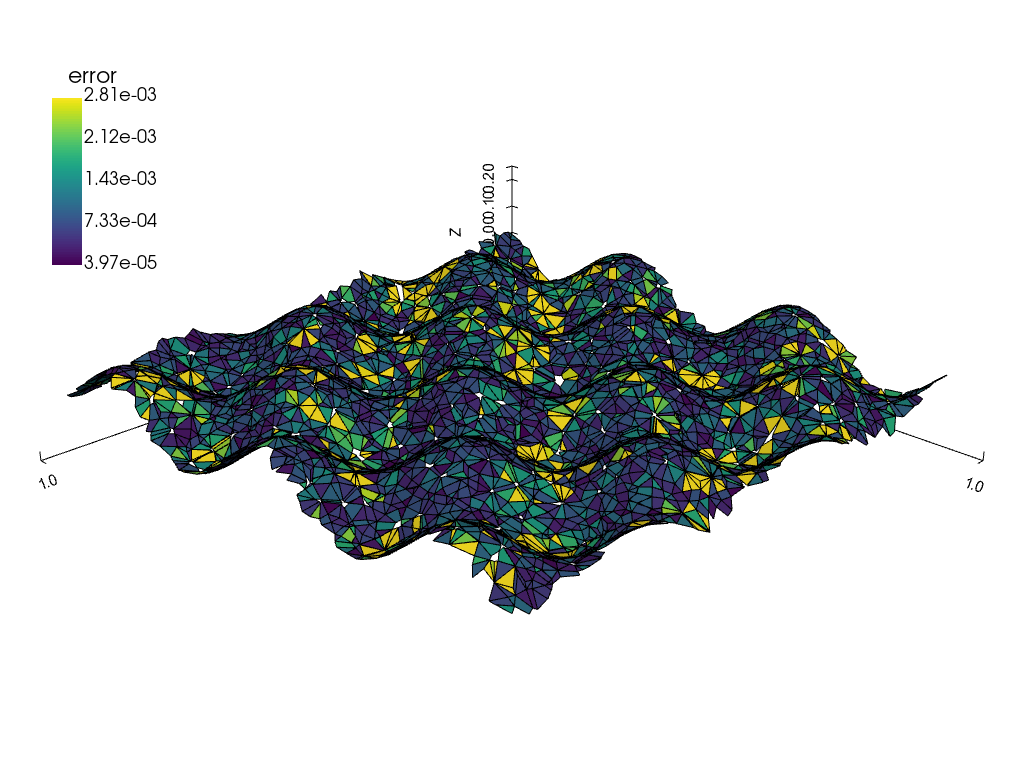}}
    \subfloat{\includegraphics[width=.33\textwidth]{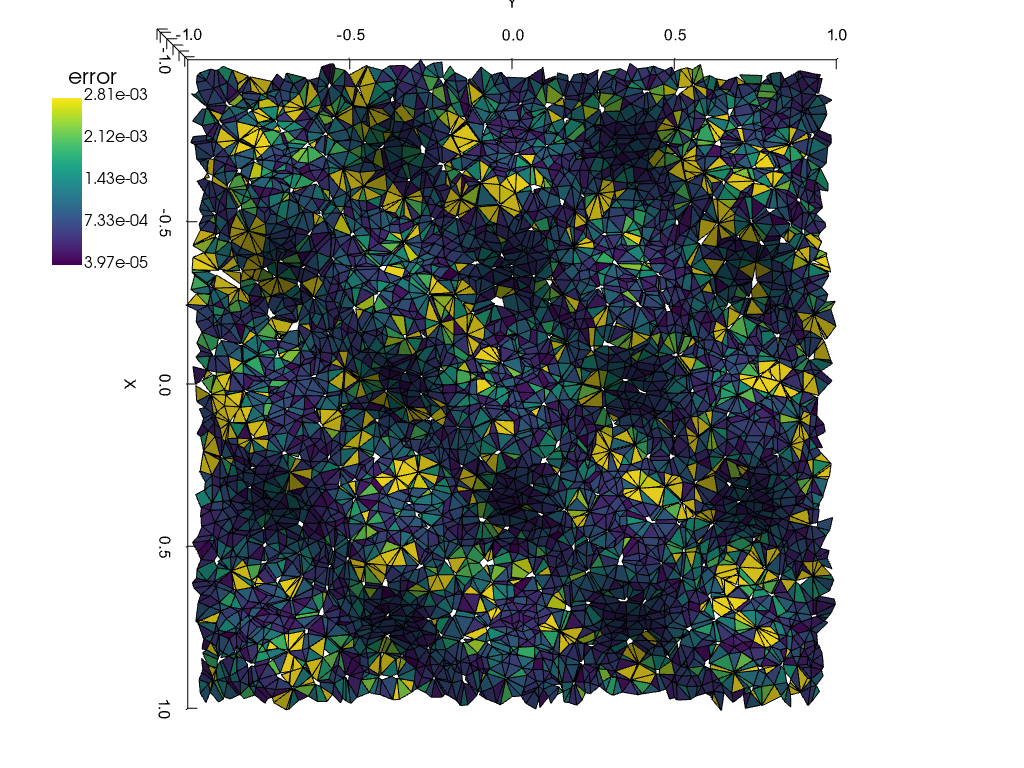}}
    \subfloat{\includegraphics[width=.33\textwidth]{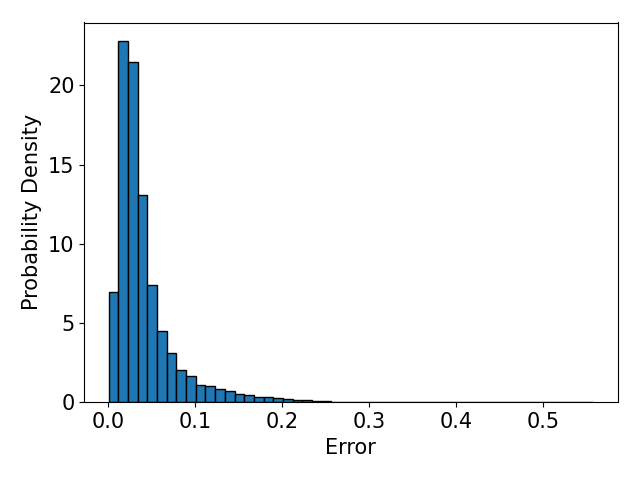}}
    \par
    \subfloat{\includegraphics[width=.33\textwidth]{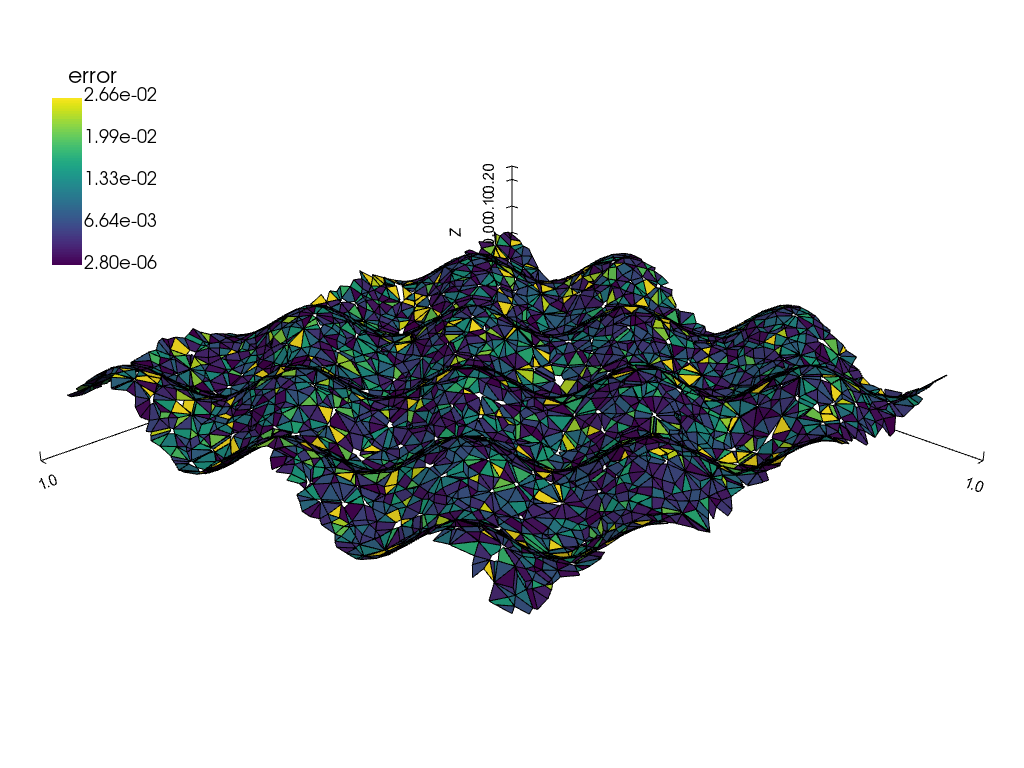}}
    \subfloat{\includegraphics[width=.33\textwidth]{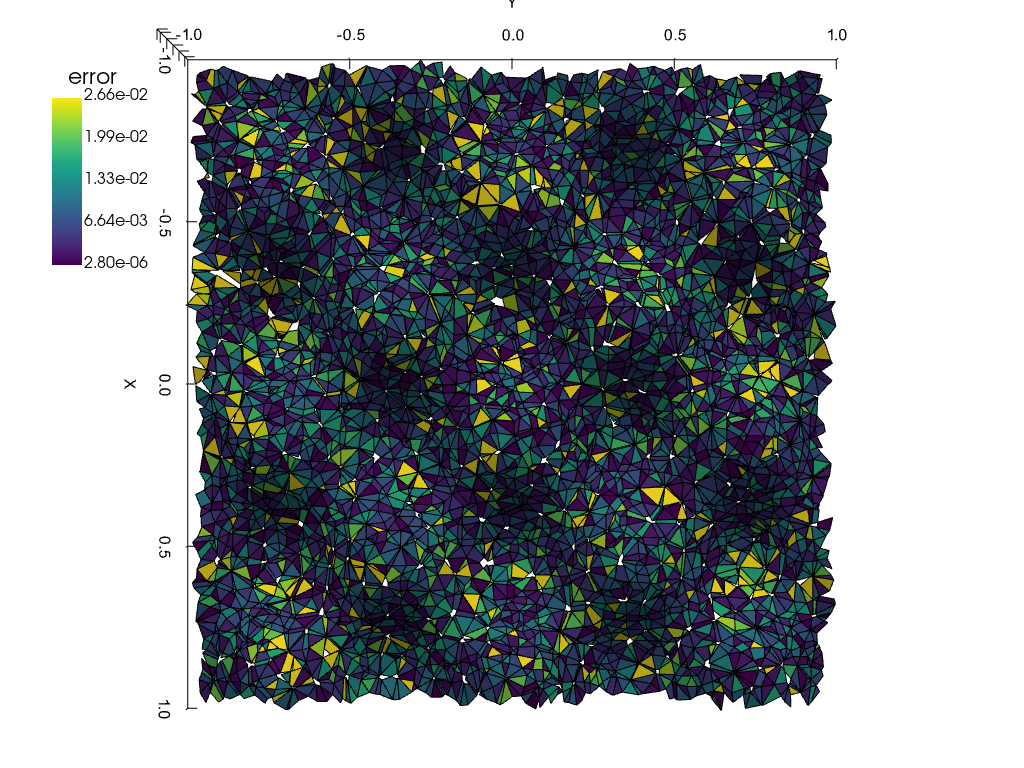}}
    \subfloat{\includegraphics[width=.33\textwidth]{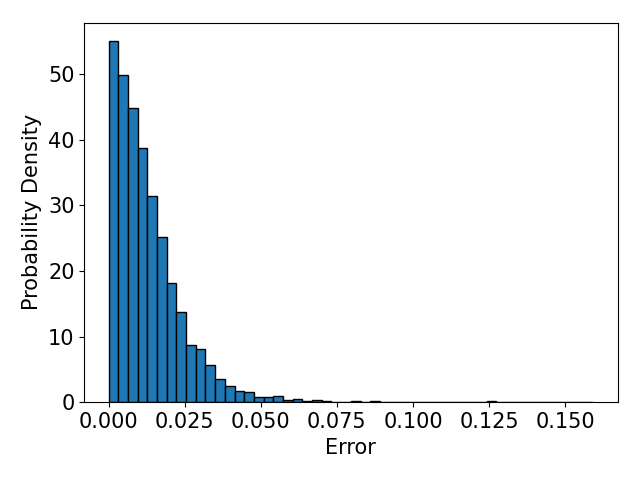}}
    \par
    \caption{Error plotted on the test surface (left and middle) and its histogram (right) of $n_{error}$, $H_{error}$, $M_{error}$, $A_{error}$ (top to bottom), the visualized range in 3D surface is indicated in the Table \ref{tab:medi_errors}}
    \label{fig:pred_hist}
\end{figure}

In order to assess the model generalization, we perform a prediction on a 3D sinusoidal surface as shown in Fig \ref{fig:test_surface}. The surface is defined by $ f(x)=0.1\sin 9x \cos 9y$. The space surrounding the surface was voxelized (generation of a tetrahedral mesh) with a constant discretization by $\Delta =0.05$. We then generate graphs in each of those voxels. Each voxel contains the center point of a graph. Its neighbor voxels contain other nodes of the same graph. This means that a voxel can be a container of a center point as well as a neighbor point. The maximum absolute non-dimensionalized curvature of the test surface is $\lvert H \rvert_{max}=0.703$. This is in the range of the training data-set ($H_{training}\approx[-0.75, 0.75]$). Each interface cell where the volume fraction is in the range $0.01<\alpha<0.99$ is input to the model, while interface cells with very high or low volume fractions are filtered out since we consider these as out of the range of applicability of our model. Fig.\ref{fig:pred_hist} shows the errors of the prediction on the four variables and their histograms. The error for each variable is defined as Table.\ref{tab:medi_errors}. We see that errors don't have a clear relation to the geometry for any variable. The histograms show a unimodal shape and no abnormal behavior is observed. Table.\ref{tab:medi_errors} also show the median of the error.

\begin{figure}\centering
    \subfloat{\includegraphics[width=.30\linewidth]{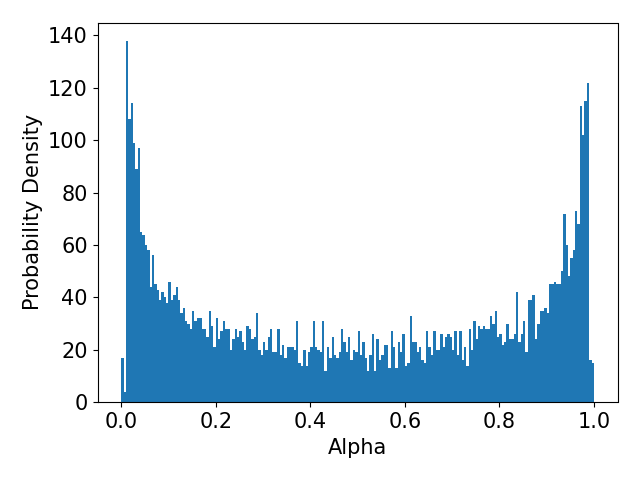}}
    \subfloat{\includegraphics[width=.30\linewidth]{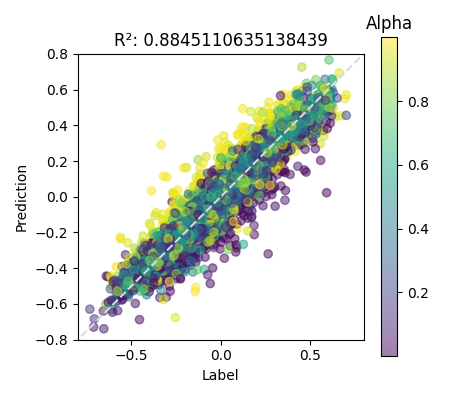}}
    \subfloat{\includegraphics[width=.30\linewidth]{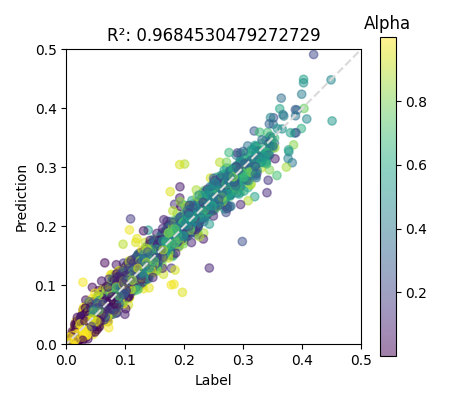}}
    \caption{Histogram of the volume fraction $\alpha$ on the test surface and $R^2$ score on the curvature (middle) and area (right) prediction colored by $\alpha$}
    \label{fig:alpha_curv}
\end{figure}

This test surface doesn't have a uniform distribution about $\alpha$ as Fig.\ref{fig:alpha_curv} shows. More samples are found at $\alpha$ close to $0$ and $1$ (hereby let us call them ``marginal $\alpha$"). This can have an impact on the prediction result. In Fig.\ref{fig:alpha_curv} the prediction is compared to the label colored by $\alpha$. The marginal $\alpha$ cases have clearly a larger error. For a small marginal $\alpha$ case the model tend to predict smaller than the label, while it tends to predict larger to a large marginal $\alpha$. This influence of $\alpha$ has not been observed in the test case in the training where $\alpha$ was quasi-uniform. It implies the limited capacity of the model on the marginal $\alpha$ and it should be taken into consideration in case of implementation in a real CFD solver and performing on real use cases.

\begin{table}[H]
\centering
\caption{Median of the error for the entire surface}
\label{tab:medi_errors}
\begin{tabular}[t]{lccc}
\hline
Predicted variable&value\\
\hline
$n_{error}=\norm{{\mathbf{n}_{label}-\mathbf{n}_{pred}}}$&$6.05e\mbox{-}2$&$[min(n_{error}), max(0.3n_{error})]$\\
$H_{error}=\abs{H_{label}-H_{pred}}$&$4.32e\mbox{-}2$&$[min(H_{error}), max(0.2H_{error})]$\\
$M_{error}=\norm{M_{label}-M_{pred}}$&$1.84e\mbox{-}2$&$[min(M_{error}), max(0.1M_{error})]$\\
$A_{error}=\abs{A_{label}-A_{pred}}$&$6.84e\mbox{-}3$&$[min(A_{error}), max(0.2A_{error})]$\\
\hline
\end{tabular}
\end{table}%

%% file: sections/conclusion-eccomas.tex
\section{Conclusions and perspectives}
In this work, we proposed a machine learning-based framework to model interface reconstruction. To do so, we generated a synthetic dataset built with paraboloid surfaces discretized on unstructured meshes. Then we introduced a GNN based method to compute the interface reconstruction on general unstructured meshes. With the generalization test, we validated our new method. We demonstrated that 1/ GNNs could be an alternative to the conventional surface reconstruction methods, 2/ GNNs could be an effective reconstruction approach to unstructured grids. This can make our method particularly interesting in an industrial context. Further work includes 1/ implementation of the model in a CFD solver, 2/ assessment of time performance in a real CFD simulation, 3/ treatment of boundary conditions.

%% file: mybib.bib
@article{harvie2006,
title = {An analysis of parasitic current generation in Volume of Fluid simulations},
journal = {Applied Mathematical Modelling},
volume = {30},
number = {10},
pages = {1056-1066},
year = {2006},
note = {Special issue of the 12th Biennial Computational Techniques and Applications Conference and Workshops (CTAC-2004) held at The University of Melbourne, Australia, from 27th September to 1st October 2004},
issn = {0307-904X},
doi = {https://doi.org/10.1016/j.apm.2005.08.015},
url = {https://www.sciencedirect.com/science/article/pii/S0307904X05001666},
author = {D.J.E. Harvie and M.R. Davidson and M. Rudman}
}

@article{qi2019computing,
  title={Computing curvature for volume of fluid methods using machine learning},
  author={Qi, Yinghe and Lu, Jiacai and Scardovelli, Ruben and Zaleski, St{\'e}phane and Tryggvason, Gr{\'e}tar},
  journal={Journal of Computational Physics},
  volume={377},
  pages={155--161},
  year={2019},
  publisher={Elsevier}
}

@article{patel2019computing,
  title={Computing interface curvature from volume fractions: A machine learning approach},
  author={Patel, HV and Panda, A and Kuipers, JAM and Peters, EAJF},
  journal={Computers \& Fluids},
  volume={193},
  pages={104263},
  year={2019},
  publisher={Elsevier}
}

@inproceedings{kothe1996volume,
  title={Volume tracking of interfaces having surface tension in two and three dimensions},
  author={Kothe, Douglas and Rider, W and Mosso, Stewart and Brock, J and Hochstein, John},
  booktitle={34th aerospace sciences meeting and exhibit},
  pages={859},
  year={1996}
}

@inproceedings{williams2002numerical,
  title={Numerical methods for tracking interfaces with surface tension in 3-D mold filling processes},
  author={Williams, Matthew W and Kothe, Doug and Korzekwa, Deniece and Tubesing, Phil},
  booktitle={Fluids Engineering Division Summer Meeting},
  volume={36150},
  pages={751--759},
  year={2002}
}

@article{williams1998accuracy,
  title={Accuracy and convergence of continuum surface tension models},
  author={Williams, MW and Kothe, DB and Puckett, EG},
  journal={Fluid dynamics at interfaces},
  pages={294--305},
  year={1998},
  publisher={Cambridge University Press Cambridge}
}

@article{cummins2005estimating,
  title={Estimating curvature from volume fractions},
  author={Cummins, Sharen J and Francois, Marianne M and Kothe, Douglas B},
  journal={Computers \& structures},
  volume={83},
  number={6-7},
  pages={425--434},
  year={2005},
  publisher={Elsevier}
}

@article{lorstad2004assessment,
  title={Assessment of volume of fluid and immersed boundary methods for droplet computations},
  author={L{\"o}rstad, Daniel and Francois, Marianne and Shyy, Wei and Fuchs, Laszlo},
  journal={International journal for numerical methods in fluids},
  volume={46},
  number={2},
  pages={109--125},
  year={2004},
  publisher={Wiley Online Library}
}

@inproceedings{hamilton2017inductive,
  title={Inductive representation learning on large graphs},
  author={Hamilton, William L and Ying, Rex and Leskovec, Jure},
  booktitle={Proceedings of the 31st International Conference on Neural Information Processing Systems},
  pages={1025--1035},
  year={2017}
}

@article{svyetlichnyy2018neural,
  title={Neural networks for determining the vector normal to the surface in CFD, LBM and CA applications},
  author={Svyetlichnyy, Dmytro},
  journal={International Journal of Numerical Methods for Heat \& Fluid Flow},
  year={2018},
  publisher={Emerald Publishing Limited}
}

@article{wu2020comprehensive,
  title={A comprehensive survey on graph neural networks},
  author={Wu, Zonghan and Pan, Shirui and Chen, Fengwen and Long, Guodong and Zhang, Chengqi and Philip, S Yu},
  journal={IEEE transactions on neural networks and learning systems},
  volume={32},
  number={1},
  pages={4--24},
  year={2020},
  publisher={IEEE}
}

@article{zhou2020graph,
  title={Graph neural networks: A review of methods and applications},
  author={Zhou, Jie and Cui, Ganqu and Hu, Shengding and Zhang, Zhengyan and Yang, Cheng and Liu, Zhiyuan and Wang, Lifeng and Li, Changcheng and Sun, Maosong},
  journal={AI Open},
  volume={1},
  pages={57--81},
  year={2020},
  publisher={Elsevier}
}
